\documentclass[preprint,12pt]{elsarticle}

\usepackage{amssymb}
\usepackage{hyperref}

\begin{document}

\begin{frontmatter}



\title{\textbf{Towards applications of graded Paraparticle algebras}}


\author[label1,label2]{\textbf{Konstantinos Kanakoglou}}
\author[label1]{\textbf{Alfredo Herrera-Aguilar}}

\address[label1]{Instituto de F\'{\i}sica y Matem\'{a}ticas (\textsc{Ifm}), Universidad Michoacana de San Nicol\'{a}s de Hidalgo (\textsc{Umsnh}), Edificio C-3, Cd. Univ., CP 58040, Morelia, Michoac\'{a}n, \textsc{Mexico}}
\address[label2]{School of Physics, Nuclear and Elementary Particle Physics Department, Aristotle University of Thessaloniki (\textsc{Auth}), CP 54124, Thessaloniki, \textsc{Greece}}

\begin{abstract}
An outline is sketched, of applications of the ideas and the mathematical methods
presented at the $19^{th}$ symposium of the Hellenic Nuclear Physics Society (\textsc{Hnps}) in Thessaloniki, May $2010$
\end{abstract}


\end{frontmatter}


\section{Introduction}
\label{Introduction} In \cite{KaDaHa1, KaDaHa2} we have studied
algebraic properties of the Relative Parabose algebra $P_{bf}$ and
the Relative Parafermi algebra $P_{fb}$ such as their gradings,
braided group structures, $\theta$-colored Lie structures, their
subalgebras etc. These algebras, constitute paraparticle systems
defined in terms of (parabosonic and parafermionic) generators
\footnote{or: interacting parabosonic and parafermionic degrees of
freedom, in a language more suitable for physicists.} and
(trilinear) relations. We have then proceeded in building
realizations of an arbitrary Lie Superalgebra (LS) $L=L_{0} \oplus
L_{1}$ (of either fin or infin dimension) in terms of these mixed
paraparticle algebras. Utilizing a given (graded), fin. dim., matrix
representation of $L$, we have actually constructed linear maps of the form
$\mathbb{J}: L \rightarrow gl(m/n) \subset \begin{array}{c}
                                          P_{bf}  \\
                                          P_{fb}
                                          \end{array}$
from the LS $L$ to a copy of the general linear
superalgebra $gl(m/n)$ isomorphically embedded into either $P_{bf}$ or into
$P_{fb}$. These maps have been shown to be graded Hopf algebra
homomorphisms or more generally braided group homomorphisms. From the
pure mathematics viewpoint, such maps may be considered as
generalizations of the Ado-Iwasawa theorems \footnote{which roughly
state that any f.d. Lie (or super-Lie) algebra is actually
isomorphic to a matrix subalgebra of $gl(n)$ (or a graded matrix
subalgebra of $gl(m/n)$).} for Lie and super-Lie algebras or even of
the Cayley theorem \footnote{that any fin. group is isomorphic to
some subgroup of the permutation group $S_{n}$.} of group theory.
From the viewpoint of mathematical physics, these maps generalize
-in various aspects- the usual bosonic-fermionic Jordan-Scwinger
realizations of Quantum mechanics. In \cite{KaHe1, KaHe2}, we have
further proceeded in building and studying a class of irreducible
representations \footnote{we have used in \cite{KaHe1, KaHe2} the
terminology ``Fock-like reprs." because these reprs. apparently
generalize the well known boson-fermion Fock spaces of Quantum Field
theory.} for the simplest case of the $P_{bf}^{(1,1)}$ algebra in a
single parabosonic and a single parafermionic degree of freedom (a
$4$-generator algebra).

\section{Prospects-Research objectives}
\label{prospects} The carrier spaces of the Fock--like
representations of $P_{BF}^{(1,1)}$ constitute a family parametrized
by the values of a positive integer $p$. They have the general form
{\small $ \bigoplus_{n=0}^{p} \bigoplus_{m=0}^{\infty}
\mathcal{V}_{m,n} $ } where $p$ is an arbitrary (but fixed) positive
integer. The subspaces $\mathcal{V}_{m,n}$ are 2--dim except for the
cases $m = 0$, $n = 0, p$ i.e. except the subspaces
$\mathcal{V}_{0,n}$, $\mathcal{V}_{m,0}$, $\mathcal{V}_{m,p}$ which
are 1--dim for all values of $m$ and $n$. These subspaces can be
visualized as {\small
$$
\begin{array}{cccccccc}
    \mathcal{V}_{0,0} & \mathcal{V}_{0,1} & \ldots  &  \mathcal{V}_{0,n} & \ldots & \ldots & \mathcal{V}_{0,p-1} & \mathcal{V}_{0,p}  \\
    \mathcal{V}_{1,0} & \mathcal{V}_{1,1} & \ldots & \mathcal{V}_{1,n} & \ldots &  \ldots & \mathcal{V}_{1,p-1} & \mathcal{V}_{1,p}  \\
    \vdots & \vdots & \ldots  & \vdots & \ldots & \ldots  & \vdots & \vdots     \\
    \mathcal{V}_{m,0} & \mathcal{V}_{m,1} & \ldots & \mathcal{V}_{m,n} & \mathcal{V}_{m,n+1} & \ldots & \ldots & \mathcal{V}_{m,p} \\
    \vdots & \vdots &  \ldots & \mathcal{V}_{m+1,n} & \ldots &  \ldots  & \vdots  & \vdots     \\
    \vdots & \vdots & \ldots & \vdots & \ldots & \ldots  & \vdots & \vdots
\end{array}
$$
}

Our research will focus on both aspects of Pure Mathematics
(developing or generalizing  techniques for building new
representations and studying their properties i.e. computing
characters, eigenvalues of Casimirs, formulae for the action of the
generators, irreducibility, ... etc.) and aspects of applying these
representations in constructing a realistic model for the
interaction of monochromatic radiation with a multiple level system:
\subsection{Representation-theoretical aspects:}
Our first objective is to study representation theoretic
implications and applications of the above mentioned constructions:
We intend to present an abstraction of the Fock representations
methodology, in such a way that it can be applicable to an arbitrary
algebra given in terms of generators and relations. This work has
already begun \cite{KaHe4}. Our next task, will be apply the method
in order to extend the results of \cite{KaHe1} to the case of the
$P_{fb}^{(1,1)}$ algebra as well and then to proceed in studying the
general cases of arbitrary degrees of freedom for either $P_{bf}$ or
$P_{fb}$. Combining this study with the results of \cite{KaDaHa1,
KaDaHa2} we will ``translate" the constructed paraparticle
representations in terms of an arbitrary Lie superalgebra. Finally,
we will proceed in studying and computing properties of the
constructed Lie Superalgebra representations such as computation of
characters, explicit formulae for the action of the generators,
eigenvalues for the Casimirs, reduction in reducible modules,
classification of the cyclic irreducible modules etc. This work has
also already begun, by considering the simplest case of
$P_{bf}^{(1,1)}$: In \cite{KaHe3} we are building Lie superalgebra
representations starting from an arbitrary Lie Superalgebra
possessing a $2$-dim. graded matrix representation.
\subsection{Towards the construction of a realistic model for the
interaction of mono-chromatic radiation with a multiple level
system:} Our second objective has to do with a potential physical
application of the paraparticle and LS Fock-like representations
discussed above, in the extension of the study of a well-known model
of quantum optics: the Jaynes-Cummings (JC) model \cite{JC} is a
fully quantized -and yet analytically solvable- model describing (in
its initial form) the interaction of a monochromatic electromagnetic
field with a two-level atom. Using the Fock-like modules described
above, we will attempt to proceed in a generalization of the above
model in the study of the interaction of a monochromatic parabosonic
field with a $(p+1)$-level system. The Hamiltonian for such a system
might be of the form {\small
\begin{equation}  \label{1}
\begin{array}{c}
\mathcal{H} = \mathcal{H}_{b} + \mathcal{H}_{f} + \mathcal{H}_{interact} = \omega_{b}N_{b}+\omega_{f}N_{f}+\lambda(Q^{+}+Q^{-}) = \\ \\
=\frac{\omega_{b}}{2}\{b^{+},b^{-} \}+\frac{\omega_{f}}{2}[f^{+},f^{-}]+\frac{(\omega_{f}-\omega_{b})p}{2}+\frac{\lambda}{2}\big(\{ b^{-},f^{+}\}+\{ b^{+},f^{-}\}\big)
\end{array}
\end{equation}
}
or more generally:
{\small
\begin{equation}  \label{2}
\begin{array}{c}
\mathcal{H} = \mathcal{H}_{b} + \mathcal{H}_{f} + \mathcal{H}^{'}_{interact} = \omega_{b}N_{b}+\omega_{f}N_{f}+\mathcal{H}^{'}_{interact} = \\ \\
=\frac{\omega_{b}}{2}\{b^{+},b^{-} \}+\frac{\omega_{f}}{2}[f^{+},f^{-}]+\frac{(\omega_{f}-\omega_{b})p}{2}+\lambda_{1}b^{-}f^{+}+\lambda_{2}f^{+}b^{-}+ \lambda_{3}b^{+}f^{-}+\lambda_{4}f^{-}b^{+}
\end{array}
\end{equation}
} where $\omega_{b}$ stands for the energy of any paraboson field
quanta (this generalizes the photon, represented by the Weyl algebra
part of the usual JC-model), $\omega_{f}$ for the energy gap between
the subspaces $\mathcal{V}_{m,n}$ and $\mathcal{V}_{m,n+1}$ (this
generalizes the two-level atom, represented by the $su(2)$
generators of the usual JC-model) \footnote{actually $\omega_{b}$
and $\omega_{f}$ might be some functions of $m$ or $n$ or both.} and
$\lambda$ or $\lambda_{i}$, $(i=1,..,4)$ suitably chosen coupling
constants. The $\mathcal{H}_{b} + \mathcal{H}_{f}$ part of the above
Hamiltonian represents the ``field'' and the ``atom'' respectively,
while the $\mathcal{H}_{interact} = \lambda(Q^{+}+Q^{-})$,
$\mathcal{H}^{'}_{interact} =
\lambda_{1}b^{-}f^{+}+\lambda_{2}f^{+}b^{-}+
\lambda_{3}b^{+}f^{-}+\lambda_{4}f^{-}b^{+}$ terms represent the
``field-atom'' interactions causing transitions from any
$\mathcal{V}_{m,n}$ subspace to the subspace $\mathcal{V}_{m-1,n+1}
\oplus \mathcal{V}_{m+1,n-1}$ (absorptions and emissions of
radiation). The Fock-like representations, the formulas for the
action of the generators and the corresponding carrier spaces, will
provide a full arsenal for performing actual computations in the
above conjectured Hamiltonian and for deriving expected and mean
values for desired physical quantities. A preliminary version of
these ideas, for the simplest case of $P_{bf}^{(1,1)}$ has already
appeared (see Section $5$ of \cite{KaHe1}). The spectrum generating
algebra of $\mathcal{H}$ may be considered to be either
$P_{bf}^{(1,1)}$ or $P_{fb}^{(1,1)}$ or more generally any other
mixed paraparticle algebra whose representations can be directly
deduced (see \cite{KaHe4} for details) from those of
$P_{bf}^{(1,1)}$ or $P_{fb}^{(1,1)}$. In this way, we will actually
construct a family of exactly solvable, quantum mechanical models,
whose properties will be studied quantitatively (computation of
energy levels, eigenfunctions, rates of transitions between states,
etc) and directly compared with theoretical and experimental
results. Last, but not least, it is expected that the study of such
models will provide us with deep inside into the process of
Quantization itself: We will be able to proceed in direct comparison
between mainstream quantization methods of Quantum Mechanics and the
idea of Algebraic (or Statistical)  Quantization (using Hamiltonians
which contain no explicit dynamical interaction terms but including
the interaction implicitly into the relations of the spectrum
generating algebra itself) as this is outlined in works such as
\cite{Pal12}.

\section{Acknowledgements}

{\small KK would like to thank the whole staff of \textsc{Ifm},
\textsc{Umsnh} for providing a challenging and stimulating
atmosphere while preparing this article. His work was supported by
the research project \textsc{Conacyt}/No. J60060. AHA was supported
by \textsc{Cic} 4.16 and \textsc{Conacyt}/No. J60060; he is also
grateful to \textsc{Sni}. }





\bibliographystyle{elsarticle-num}
\bibliography{<your-bib-database>}



\end{document}